\documentclass[manuscript,nonacm]{acmart}

\AtBeginDocument{\providecommand\BibTeX{{\normalfont B\kern-0.5em{\scshape i\kern-0.25em b}\kern-0.8em\TeX}}}

\newif\ifanonymous
\anonymousfalse
\newcommand{\anonymize}[2]{\ifanonymous #2\else #1\fi}
\newcommand{\anonymizeWithBlanks}[1]{\ifanonymous $\blacksquare\blacksquare\blacksquare\blacksquare\blacksquare$\else#1\fi}
\usepackage{xspace}
\newcommand{\Iltis}{\textsc{Iltis}\xspace}

\usepackage{natbib}
\usepackage{svg}
\usepackage{csquotes}
\usepackage{tabularx}

\begin{document}

\title{Exploring Error Types in Formal Languages Among Students of Upper Secondary Education}

\author{Marko Schmellenkamp}
\authornote{The first two authors contributed equally to this work.}
\orcid{0000-0003-3966-6590}
\affiliation{\institution{Ruhr University Bochum}
  \streetaddress{Universitätsstraße 150}
  \city{Bochum}
  \country{Germany}
}
\email{marko.schmellenkamp@rub.de}

\author{Dennis Stanglmair}
\authornotemark[1]
\affiliation{\institution{TU Munich, TUM School of Social Sciences and Technology}
  \streetaddress{Arcisstraße 21}
  \city{Munich}
  \country{Germany}
}
\email{}

\author{Tilman Michaeli}
\orcid{0000-0002-5453-8581}
\affiliation{\institution{TU Munich, TUM School of Social Sciences and Technology}
  \streetaddress{Arcisstraße 21}
  \city{Munich}
  \country{Germany}
}
\email{tilman.michaeli@tum.de}

\author{Thomas Zeume}
\orcid{0000-0002-5186-7507}
\affiliation{\institution{Ruhr University Bochum}
  \streetaddress{Universitätsstraße 150}
  \city{Bochum}
  \country{Germany}
}
\email{thomas.zeume@rub.de}

\begin{abstract}
    Foundations of formal languages, as subfield of theoretical computer science, are part of typical upper secondary education curricula. There is very little research on the potential difficulties that students at this level have with this subject. In this paper, we report on an exploratory study of errors in formal languages among  upper secondary education students. We collect the data by posing exercises in an intelligent tutoring system and analyzing student input. Our results suggest a) instances of non-functional understanding of concepts such as the empty word or a grammar as a substitution system; b) strategic problems such as lack of foresight when deriving a word or confounding formal specifications with real-world knowledge on certain aspects; and c) various syntactic problems.
    These findings can serve as a starting point for a broader understanding of how and why students struggle with this topic.
\end{abstract}



\maketitle

\section{Introduction}

The foundations of formal languages are a cornerstone of many areas of computer science programming languages and formal verification. Introductions to formal languages are therefore an integral part of standard undergraduate and K-12 curricula \cite{ACM2013, Webb17}. Experience suggests that learning foundations of formal languages, as part of the theory of computing education, is difficult for many students, likely because it requires using mathematical thinking and techniques. 

The basis for addressing students' difficulties with this topic is a solid understanding of what exactly is difficult for students and what kind of non-feasible conceptions they have. With this knowledge, material can be designed that takes into account students' conceptions and/or directly targets common errors. Such insights are also useful in the design of intelligent tutoring systems, as they can form the basis for designing more helpful feedback on students' submissions. Thus, understanding common difficulties is essential for improving our formal language teaching.

Educational research in computer science has largely focused on programming education, where extensive research has been conducted on difficulties and misconceptions (see, e.\,g., \cite{Qian17}), including studies aimed at (higher) secondary education (see, e.\,g., \cite{SwidanHS18}). Unfortunately, there is almost no educational research on the foundations of formal languages.

In this paper, we start an exploration of students' understanding of formal languages, with a focus on context-free language representations and students in secondary education.

\begin{itemize}
    \item[\textbf{RQ}] What errors do upper secondary students make in exercises on formal languages?
\end{itemize}

To address this question, we explore errors made by students in an introductory course on formal languages, in particular interpreting and constructing formal grammars and translating them to other representations. To collect errors, we developed interactive material in \anonymize{the \Iltis system \cite{GeckLPSVZ18, SchmellenkampVZ24}}{an intelligent tutoring system} tailored to the curriculum of year twelve in a \anonymize{German}{} grammar school and accompanied its use in class.
By means of a qualitative content analysis, we then categorized the errors found.
We also discuss our findings with the teacher of the course and gain insights into the students' use of the system in class.

Our study is intended to facilitate and guide future research in identifying and addressing common errors in formal languages.

\section{Related Work}

Computer science education is an active area of research covering a wide range of topics including curriculum design, errors and misconceptions, etc. Most research in this area has focused on programming education, while theory of computing education has been largely neglected.

In the following we first review the little existing research on theory of computing education (in Section \ref{section:formal-method-education}) with a focus on formal languages, and then describe potential insights for theory of computing education that can be drawn from related fields (in Section \ref{section:related-fields}), in particular from mathematics education and programming education.

All of the research on theoretical computer science education mentioned in Section \ref{section:formal-method-education} is conducted for higher education. For the well-researched field of programming, also (upper) secondary education is targeted (see, e.\,g.,\cite{SwidanHS18}), for many other areas, in particular for theoretical computer science, this is still missing.

In summary, there is great potential but only very little literature on understanding students' errors in theoretical computer science, in particular when secondary education is concerned.

\subsection{Theoretical Computer Science Education}
\label{section:formal-method-education}
It has been argued that theoretical computer science is an essential part of curricula for computer science as well as software engineering degree programs \cite{ACM2013}. From a research point of view, several studies focus on developing a curriculum and general methodology for teaching theoretical foundations \cite{BushmelevaB17,TavolatoV12,SpichkovaZ16} on a higher education level. Other studies focus on experience reports in delivering such courses \cite{Scheurer00,IshikawaYT15}, particularly in engaging students in the study of formal languages and theory of computing \cite{Sigman07}.

Regarding specific content, however, there is only very little previous work: As a rare example, \citeauthor{GalEzerT16} \cite{GalEzerT16} have done research on problems of students with computational reductions. And for automata theory, it has been found that simulating automata for specific words and therefore testing them for single inputs, is a sensible first step for students to reason about these automata \cite{Souza15, MorazanA14}.

Teaching theoretical computer science is supported by a variety of support systems covering many aspects including logic, formal languages, proof techniques, etc. For formal languages, well-known systems include \emph{AutomataTutor} \cite{AlurAGKV13,AntoniKAGV2015,AntoniHKRW2020}, \emph{Exorciser} \cite{Tscherter04}, \emph{FLACI} \cite{HielscherW19}, \emph{Iltis} \cite{GeckLPSVZ18, SchmellenkampVZ24}, \emph{JFLAP} \cite{Rodger1999,GramondR1999}, and \emph{RACSO} \cite{CreusG14}.
Besides supporting students, these systems also have great potential for collecting educational data to gain insights into students' thinking. For example, the \emph{Iltis} system has been used to detect common errors in modeling with formulas \cite{SchmellenkampLZ23}.

\subsection{Insights from Related Fields}
\label{section:related-fields}
While there has been only few research on theoretical computer science education, insight may be gained from several related fields such as mathematics education and programming education.

Building abstractions and then working on these abstractions with a well-defined set of techniques is an important aspect of theoretical computer science. One such abstraction are formal grammars, which come with a set of techniques for transforming grammars into other representations, testing whether a word can be derived, etc. In general, building abstractions was shown to be difficult for many students over several domains, including programming education \cite{GinatB17} and working with algebra in mathematics education \cite{Sfard95}. To support students in the process of building and working with abstractions, methodological concepts have been devised, one of the best known being the concrete-to-representational-to-abstract (CRA) model \cite{WitzelRS08}.

Formal grammars can be interpreted as rule-based (substitution) system, which may explain certain errors or conceptions of students. Misconceptions are sometimes not caused by the topic at hand but by difficulties of applying rules of some rule-based systems \cite{YoungO81}. This line of research is backed up by many results, for instance also for working with functional programs which also can be seen as rule-based systems \cite{DavidLMC93}. 

Another aspect of formal grammars that may add complexity is recursion. Recursive non-terminals are necessary for grammars specifying infinite languages. In programming, the concept of recursion has been proven to be difficult for many students (see e.\,g. the survey \cite{McCauleyGFM15}). While recursive rules in formal grammars are not the same as recursive calls in programs, these concepts are close enough to ask whether conceptional problems transfer.

In a literature survey of programming education by \citeauthor{Qian17} \cite{Qian17}, a categorization of difficulties in three knowledge domains has been proposed: difficulties in syntactical, conceptual, and strategic knowledge. For example, mismatching parentheses is a typical difficulty in syntactical knowledge. Examples of difficulties in conceptual knowledge include the non-awareness of students that instructions are executed in the given order or that a variable can hold only one value at a time; however for instance a failure in using a variable as a counter variable of some sort is classified as difficulty in strategic knowledge. For misconceptions in programming, also the awareness of the teachers about these misconceptions (see \cite{QianHYGL20}) and possible support by specially designed tools has been extensively studied (see e.\,g. \cite{MohammedS24}).

\section{Methodology}

The aim of our study is to to address the research gap concerning errors and conceptions in formal languages education in K-12. We sketch our methodology below.

\subsection{Setting and Context}

We conducted our study in an elective  computer science class in year 12 of a grammar school in \anonymizeWithBlanks{Germany}. The students were in their second-last year of secondary education and the class consisted of 13 male and 3 female students aged between 16 and 17.
The students were taught formal languages for about 10 weeks with 3 lessons à 45 minutes per week. 
According to the teacher's assessment, the class was very heterogeneous in terms of performance and motivation while the overall level of achievement was average.

During our study, the following concepts were taught in this course:
\begin{itemize}
    \item Foundations of formal languages: formalization of real-world settings as formal languages; differentiation of syntax and semantics of formal languages
    \item Formal grammars: interpreting context-free grammars; modeling formal languages with context-free grammars
\item Alternative language representations: Syntax diagrams; grammars in extended Backus-Naur form (EBNF); conversions between these representations
\item Finite automata: interpreting deterministic finite automata (DFA); constructing DFA for regular languages
\end{itemize}

In the classroom, the teacher used a conventional textbook, analogue exercise sheets, and additional interactive material in \anonymize{the intelligent tutoring system \Iltis}{an intelligent tutoring system} provided by us. The students worked on the interactive material in the classroom as well as individually for homework.

In the material, each concept covered (alphabets, words, formal grammars, finite automata) was introduced by interleaving definitions and examples with interactive exercises in the \anonymize{\Iltis}{tutoring} system. This is a well-established technique, see, e.\,g.~\cite{MohammedS24}. Students could then practice the concept by solving multiple interactive exercises. All exercises were designed to be closely aligned with the curriculum. For all interactive exercises, the students had an unlimited number of attempts, and after each attempt, the intelligent tutoring system provided feedback on the correctness and, where appropriate, explanations of why attempts were incorrect.
The interactive exercises were of one of the following types: specifying words of formal languages (satisfying some constraints); constructing formal grammars for regular languages; deriving words of grammars step-by-step; constructing finite state automata for regular languages; and multiple choice questions.

\subsection{Data Collection and Analysis}

Students' inputs for the interactive exercises were anonymously logged.
There were 23 exercises for specifying a word with 757 attempts in total (343 of them being correct), 7 exercises for constructing a grammar with 240 attempts in total (37 of them being correct), and 7 exercises for constructing a derivation with 1198 attempts in total (777 of them being correct). Note that for the exercises on constructing a derivation, each derivation step is considered an attempt.
For technical reasons, no logging was possible for tasks on constructing automata and multiple choice questions.

For the data collected from the tutoring system, we conducted a structured qualitative content analysis according to Mayring \cite{Mayring2014}. The software MaxQDA was used for the
coding.
The use of a qualitative data analysis method allowed us to examine individual student responses in depth, identifying patterns, themes, and recurring issues.
Given the lack of existing work investigating learners' errors in the context of formal languages, we  analyzed students' inputs inductively. Inputs could be assigned to multiple categories if they exhibited characteristics of more than one error. At regular intervals in the categorization process, the category system was refined and condensed. This involved merging similar categories observed at different task types, eliminating redundancy, and prioritizing the most salient themes and patterns observed in the data. 

The categorization was mainly carried out by one of the co-authors. At multiple times during this process, the current state of the category system was discussed within the research team (with experiences in teaching formal languages at a university level) in order to agree on the coding scheme,  categorization, and category descriptions. 

To provide a comprehensive overview of the inductively derived error categories, we structured them according to the system proposed by \citeauthor{Qian17} \cite{Qian17}, which categorizes errors into three knowledge domains: syntactical, conceptual, and strategic knowledge. This appeared to be very suitable, as all three knowledge domains play an important role in the field of formal languages.

After analyzing the data, we interviewed the teacher who conducted all the lessons to discuss our findings and gain additional insight from her perspective. This collaboration allowed us to contextualize and back our findings within the specific classroom environment.

\newcommand{\resulttable}{
    \begin{table}
        \centering
        \begin{tabularx}{\linewidth}{lXlll}
        \toprule
        \multicolumn{2}{l}{\textbf{Categories}}&  \textbf{Exercise type} & \textbf{\# tasks} & \textbf{\# attempts} \\
        &&&\textbf{err. / total}&\textbf{err. / total}\\\midrule
        \multicolumn{5}{l}{\textbf{S Syntactical difficulties}}\\\midrule
        &S1 Notation and formatting    
                & construct derivation  & 3 / 7& 43 / 548\\
        &       & construct grammar     & 5 / 7& 59 / 203\\\midrule
        &S2 Application and combination of productions    
                & construct derivation  & 5 / 7& 61 / 1131\\
        &       & construct grammar     & 1 / 7& 6 / 86\\\midrule
        \multicolumn{5}{l}{\textbf{C Conceptual difficulties}}\\\midrule
        &C1 Concept of the empty word 
                & specify word          & 4 / 23& 19 / 402\\
        &       & construct derivation  &  4 / 7& 24 / 736\\\midrule
        &C2 Substitution concept of productions    
                & construct derivation  & 6 / 7& 75 / 1179\\
        &       & construct grammar     & 1 / 7& 4 / 86\\\midrule
        &C3  Recursion    
                & construct derivation  & 2 / 7& 6 / 452\\
        &       & construct grammar     & 2 / 7& 17 / 54\\\midrule
        &C4 Differentiation of termini
                & specify word          & 5 / 23& 38 / 532\\
        &       & construct grammar     & 5 / 7& 9 / 203\\\midrule
        \multicolumn{5}{l}{\textbf{P Strategic difficulties}}\\\midrule
        &P1 Interpreting natural language problems    
                & specify word          & 6 / 23& 254 / 668\\
        &       & construct derivation  & 4 / 7& 30 / 887\\
        &       & construct grammar     & 4 / 7& 13 / 181\\\midrule
        &P2 Transitioning between language representations    
                & construct grammar     & 2 / 7& 37 / 80\\\midrule
        &P3 Planning derivations    
                & construct derivation  & 1 / 7& 23 / 196\\\bottomrule
\end{tabularx}
        \caption{An overview of the categories of errors observed and the task types in which they occurred. The third column shows in how many exercises of the respective type the error was observed and the total number of exercises of that type. The fourth column is with respect to exercises where this error type was observed. It shows the number of student attempts with this error and the total number of student attempts for these exercises.}
        \label{table:quant}
    \end{table}
}

\section{Results}
\label{sec:results}
In this section, we present the category system according to the three knowledge domains (syntactical, conceptual, and strategic knowledge) that results from our analysis. For each category, we (1) describe the error category, (2) illustrate errors in the category for different exercise types with examples from our data, and (3) discuss further insights from our interview with the teacher. Table \ref{table:quant} provides an overview of the categories.
\anonymize{All exercises were posed in German, for giving examples here we translate them into English.}{}

\resulttable

\subsection{Syntactical Difficulties}
In this section, we report errors in syntactical knowledge, i.\,e. in reading and writing formal representations.

\paragraph*{S1: Notation and formatting errors}
This category describes errors in the construction of grammars and the derivation of words. In the context of constructing grammars, this encompasses incorrectly labeling terminals and non-terminals (as per convention, non-terminals are to be enclosed by <\dots> and terminals by \textquotesingle\dots\textquotesingle), omitting the arrow in productions, and labeling the pipe symbol as terminal symbol, though the pipe is used in grammars for separating productions for the same left-hand side. Additionally, specifying productions with multiple non-terminals on the left-hand side or a missing right side is included in this category.
With regard to the derivation of words, this category encompasses syntax errors in the single sentential forms of each step, for example the sentential forms including line breaks or arrows.

It is challenging to differentiate between careless errors and actual deficiencies in syntactical knowledge. This observation was also made by the teacher, who noted that such errors were prevalent throughout the entire process of working on the topic.

\paragraph*{S2: Incorrect application and combination of productions}
The error category S2 encompasses two related types of syntactical errors: a) syntactical errors in the notation of derivation steps, and b) syntactical errors in constructing sequences of productions in a grammar.

The application of a production $X\to\alpha$ in a derivation step replaces (exactly) one occurrence of $X$ in the current sentential form with $\alpha$, while the remaining part of the sentential form remains unchanged. Errors of type a) indicate that students fail to retain all of the non-changed parts of the sentential form, resulting in a syntactically invalid derivation step. This category also encompasses instances where students (correctly) perform multiple derivation steps simultaneously, despite the assignment only allowing one derivation step at a time. The teacher reported that students continued to do that, even in later assessments.

The following example illustrates a combination of the aspects mentioned above: Aiming to derive the word \texttt{L001} from a grammar describing room number of a school building, a student made the following derivation step:
\[\texttt{L<Level><Room>}\Rightarrow\texttt{0<Digit><DigitWithoutZero>}\]
The student presumably wanted to simultaneously apply productions for both non-terminals of the sentential form on the left-hand side. The respective productions of the given grammar were:
\begin{align*}
    \texttt{<Level>}&\rightarrow\texttt{\textquotesingle0\textquotesingle}\mid\texttt{\textquotesingle1\textquotesingle}\mid\texttt{\textquotesingle2\textquotesingle} &
    \texttt{<Room>}&\rightarrow\texttt{<Digit><DigitWithoutZero>}
\end{align*}
However, the student failed to include the already derived \texttt{L} at the start of the second sentential form.

Errors of type b) comprise students chaining separate productions into one (syntactically incorrect) rule using multiple arrow symbols.
This can be observed for example in the following attempt of a student constructing a grammar describing blood types:
\[\texttt{<Bloodgroup>}\rightarrow\texttt{\textquotesingle A\textquotesingle <Factor>}\rightarrow\texttt{\textquotesingle+\textquotesingle}\mid\texttt{\textquotesingle-\textquotesingle<Group>}\rightarrow\texttt{\textquotesingle R\textquotesingle \textquotesingle h\textquotesingle}\]
When focusing on one arrow at a time, we see correct ideas such that each blood type consists of a letter (e.\,g., \texttt{A}), the string \texttt{Rh}, and a factor (\texttt{+} or \texttt{-}). However, the formatting and also the omission of the remaining options given in the assignment (e.\,g., \texttt{B} and \texttt{AB}) make this grammar attempt resemble a derivation.
In fact, the concept of a derivation was introduced directly below this exercise, including a definition and an example, which may have influenced students.

\subsection{Conceptional Difficulties}
Conceptual difficulties refer to problems in understanding and applying concepts of formal languages.

\paragraph{C1: Concept of the empty word is unclear}
The errors classified in category C1 concern the notion of the \emph{empty word}, denoted by $\varepsilon$ in the course studied. 

Students frequently included $\varepsilon$ within words and in particular within sentential forms in derivation steps.
For example, by applying the production \[\texttt{<Cheese>} \rightarrow \texttt{\textquotesingle Cheddar\textquotesingle}\mid\texttt{\textquotesingle Gouda\textquotesingle}\mid\varepsilon\] from a grammar for specifying sandwiches, students derived $\texttt{SesameHamTomato}\varepsilon$ from $\texttt{SesameHamTomato<Cheese>}$. However, $\texttt{SesameHamTomato}\varepsilon$ is not an (atomic) word but rather a concatenation of two words since $\varepsilon$ is itself a word rather than a terminal symbol.
Students may have included $\varepsilon$ in the derived word because they were unaware that $\varepsilon$ actually represents the empty word and therefore should be omitted. They may also not have realized that $\varepsilon$ is a word instead of a symbol from the alphabet.
Alternatively, they may have made this error in an attempt of being overly verbose in making each derivation step, or because they did not know that only atomic words should be given. The latter two reasons can be attributed to the wording of the assignments rather than to the  students' underlying conceptions.
When we discussed this finding with the teacher, we were informed that there may have been some misinformation about the concept of $\varepsilon$ in the classroom, which may have contributed to these errors.

In addition, students often struggled to determine whether $\epsilon$ was part of a formal language or not. This could be observed in tasks in which students were asked to specify words of given languages. In these instances, they incorrectly answered that the empty word was contained in the language. In addition to suggesting a conceptual issue with the concept of the empty word, another reason for this error is that in tasks for entering words, the empty word is the default option when no other input is specified.

\paragraph*{C2: Substitution concept of productions is unclear}
Category C2 encompasses errors related to applying and constructing production rules of formal grammars.
Category S2 also deals with applying productions, but while Category S2 focuses on syntactical deficiencies in sound derivation steps, this category focuses on conceptually incorrect derivation steps.

In derivation tasks, 
students frequently made derivation steps that did not correspond to any production of the given grammar.
In some cases, it was possible to infer which production the students (probably) wanted to apply, but they failed to do so. For example, when deriving words from a grammar that describes room numbers, students incorrectly applied the production \[\texttt{<Room>}\rightarrow\texttt{<Digit><DigitWithoutZero>}\]
to the sentential form \texttt{L0<Room>} and derived \texttt{L0<DigitWithoutZero>} or \texttt{L00<Room>}. In the first example, the occurrence of the non-terminal \text{<Room>} was replaced by only parts of the right-hand side of the mentioned production, while in the second example the next terminal symbol (\texttt{0}) was simply inserted without replacing the non-terminal.
The teacher attributes the errors in this area to potential errors during the initial introduction to the topic but cannot confirm this in subsequent tests.

We also include the following observation in this error category:
In formal grammars, the order in which the single productions are specified does not matter as long as the start symbol is known. Typically, grammars in educational contexts are presented roughly in a top-down fashion, specifying the productions for the start symbol first, followed by the productions of the non-terminals that are referenced by the preceding productions. In particular, with this convention the start symbol does not need to be given explicitly.
In some instances however, students entered grammars in a bottom-up fashion, placing the productions of the start symbol last. As the intelligent tutoring system used did not have an option for explicitly entering the start symbol, we do not know whether these students were aware of what the start symbol of their grammar is.

\paragraph*{C3: Difficulties with recursion}
Error category C3 describes errors in dealing with recursive non-terminals, i.\,e. non-terminals $X$ that allow for a derivation of the form $X\Rightarrow\alpha X\beta$ for any sentential forms $\alpha$ and $\beta$.
We observed two types of problems with recursion in grammars:

In tasks requiring the construction of grammars, students encountered difficulties in building a recursive non-terminal.
The following example is taken from an exercise to construct a grammar for barcodes and shows a student attempt of constructing a production allowing for an unlimited concatenation of \texttt{<White><Black>}, representing white and black stripes of the barcode. The student entered the following productions for the non-terminal \texttt{<Pair>}: \[\texttt{<Pair>}\rightarrow\texttt{<Pair>}\mid\texttt{<White><Black>}\]
A correct definition of \texttt{<Pair>} would have been: \(\texttt{<Pair>}\rightarrow\texttt{<Pair><Pair>} \mid \texttt{<White><Black>}\)

In tasks for deriving words, students frequently used a production terminating the recursion either too early or too late, which prevented them from deriving the correct word.
For example, when using the production \[\texttt{<Toping>}\rightarrow\texttt{<Toping><Toping>} \mid \texttt{\textquotesingle Pickle\textquotesingle}\mid\texttt{\textquotesingle Tomato\textquotesingle}\] to derive the word \texttt{PickleTomato} from the non-terminal \texttt{<Toping>}, students did not apply the production $\texttt{<Toping>}\rightarrow\texttt{<Toping><Toping>}$ first, but immediately used the production $\texttt{<Toping>}\rightarrow\texttt{\textquotesingle Pickle\textquotesingle}$ and thus reached a sentential form from which no successful derivation of the target word was possible anymore.

However, difficulties in selecting the proper production to continue a derivation also appeared when no recursion was involved. Errors of this type form category P3.

\paragraph*{C4: Differentiation of termini is unclear}
As a recurring issue, we noticed that students frequently conflated and misused the terms non-terminals, terminals, words, alphabets, and languages.
For example, when asked about the shortest word in a language, some students provided the shortest word over the alphabet or the alphabet itself. In some instances, terminals not included in the alphabet were used, both in grammar construction and in specifying words.

The teacher identified the conflation of different terms as a common problem, which is exacerbated by the addition of more complexity, building on these basic ideas.

\subsection{Strategic Difficulties}
In this section, we examine challenges associated with applying problem-solving strategies, with a particular emphasis on the different representations of formal languages.

\paragraph*{P1: Difficulties of interpreting natural language problems}
The error category P1 encompasses the difficulties students encounter when interpreting formal languages described in natural language. These difficulties manifested themselves in various exercises for constructing grammars, specifying words, and deriving grammars by students not adhering to all the requirements specified by the given language.

Given the contextualization of tasks to contexts familiar to students, it was observed that students relied on real-world experiences rather than the formal requirements of the given language. One illustrative example is an exercise that required students to derive words from a grammar describing room numbers of a school building. The assignment described that room numbers for computer labs are prefixed with the letter \enquote{I}. However, students frequently used \enquote{C} instead, which was used for chemistry labs in the assignment. In fact, in the students' school, computer labs were indeed labeled with the prefix \enquote{C}, which suggests that the students used their real-world knowledge rather than the information from the assignment.
This behavior was observed in numerous tasks across all task types. Inattentive reading was identified by the teacher as a potential contributing factor to such errors. Students tend to skim through task descriptions and then rely on their assumptions about how the task functions.

\paragraph*{P2: Difficulties in transitioning between language representations}
The error category P2 refers to difficulties encountered when translating between different formal language representations. Our material only contained tasks for translating from syntax diagrams to equivalent context-free grammars. We observed that some students omitted parts of syntax diagrams, particularly optional paths and alternatives. They appeared to translate a single path through the syntax diagram instead of all of them.
The teacher described the transfer between representations as a general challenge for her students.

\paragraph{P3: Difficulties in planning a derivation}
Category P3 refers to deriving a word from a grammar that has multiple similar productions for the same non-terminal.
In order to be able to derive the target word from such grammars, students have to carefully choose which of the possible productions to apply in each step. To do so, they must have a mental model of the complete derivation from the first steps onward. Some students struggled with that and chose non-appropriate productions. They performed syntactically correct derivation steps but were not able to derive the target word in the end.

\section{Discussion}

In this study, we investigated the errors that students made when working on assignments in our intelligent tutoring system. To this end, we logged their inputs and  categorized them in different error categories. We structured these categories according to the different knowledge domains used in \cite{Qian17}.

Some of our findings were anticipated. For instance, it is not surprising that some students require time to internalize the introduced formalisms. This is particularly evident in light of the fact that for many students of the course studied, this was their first encounter with strict formalisms other than numerical variables in mathematics education. The difficulty of formalisms for students can also be observed at university level.
A similar situation arises with regard to issues with recursive non-terminals. As is well-known in literature (see, e.g., the survey \cite{McCauleyGFM15}), students tend to experience difficulties with recursion in programming. It was therefore to be expected that there would be problems with recursion in grammars, too.

A number of errors found in this study are also observed in higher education settings. These include the incorrect use of the empty word, conflations of the defined terms (grammar, word, alphabet, etc.), and strategic issues when constructing a formal (context-free) grammar for a given language. These observations suggest that these aspects may be intrinsically difficult for novices, regardless of their educational stage.

Finally, we identified unexpected errors. In particular, we did not anticipate that the concept of productions as a substitution system (C2) to be a significant challenge for students. It was informative to observe some students modeling grammars in a bottom-up fashion. While this approach is uncommon for specifying grammars (in educational settings), a parallel can be drawn with programming, where each variable must be assigned a value before it can be used in further computations. Naturally, this resemblance comes to an end when recursive non-terminals are concerned, leading to the question whether students modeling in a bottom-up fashion have greater difficulty with recursive grammars.

It was also enlightening to witness an account of how unrelated content in proximity to an assignment can (probably) influence students' inputs. In the case observed in the data (S2), we hypothesize that some students who had not yet internalized the concept of a formal grammar saw an introduction to derivations while searching for hints on how to construct a grammar. Their understanding of grammars might then have been conflated with the example derivation they had encountered.

\subsection{Limitations}

We focused only on a small subset of theoretical computer science: formal languages and, in particular, formal grammars.

The strongest limitation of our study is that we collected our data from a single class.
Thus, the reported results only represent students' experiences in a single educational setting, and we cannot generalize on possible errors in other settings. We therefore refrain from making statements about the commonness of the error categories found.

Furthermore, we did not have precise control over how the teacher used the material provided in the intelligent tutoring system. In particular, we could not control know how long and in which phase of learning the students worked with the prepared material. To address the influence of how the system was used, we interviewed the teacher. To further reduce the influence of the individual teacher, each concept was introduced in a self-contained way in the material in the system. However, as seen for some errors in category S2, the material itself 
may also have influenced students.

Finally, we need to reflect critically on the categorization process. In some cases the categorization into error categories was based on an interpretation of the students' inputs. This is in particular true for grouping the single error types to the targeted knowledge domains. In fact, without further information, it is not possible to see the intention of  students in their attempts and to distinguish careless errors from errors due to non-suitable conceptions.
In order to mitigate this issue, the author team thoroughly reviewed and discussed the resulting category system.
This also points to the need for further work to identify the causes (such as underlying misconceptions) of the errors we identified.

\section{Conclusion and Future Work}

Our results broaden the picture we have about teaching formal languages, despite the limitations discussed above. By presenting categories of errors that occurred in a relatively small course, we lay the groundwork for further studies to verify and quantify the occurrence of these errors.

Our study serves as a starting point for understanding difficult aspects of teaching formal languages. It provides a basis for investigating prevalent student conceptions and the reasons behind the errors we have identified. Based on our category system, subsequent studies can explore whether the error types described in this paper manifest themselves in other settings as well.
In particular, the influence of the interactive material developed for this study can also be studied. 

Finally, studying the frequencies of the observed errors across a wide variety of settings is crucial for identifying the most relevant problems. Intelligent tutoring systems are an appropriate tool for conducting this type of quantitative study, as the data collection scales well with an increasing number of participating students. In order to automate the analysis of students’ attempts, another future perspective is to formalize how to automatically recognize these error types. This is not only helpful in conducting studies but is also the basis for individual feedback targeted to individual errors in intelligent tutoring systems. 
\begin{acks}
    We thank Antje Hanusch for using Iltis in their classroom and providing valuable feedback.
    This work was supported by the Deutsche Forschungsgemeinschaft (DFG, German Research Foundation), grant 448468041. It has benefited from Dagstuhl Seminar \href{https://www.dagstuhl.de/24251}{24251} \enquote{Teaching Support Systems for Formal Foundations of Computer Science}.
\end{acks}

\bibliographystyle{ACM-Reference-Format}
\bibliography{sources}


\begin{thebibliography}{33}


\ifx \showCODEN    \undefined \def \showCODEN     #1{\unskip}     \fi
\ifx \showDOI      \undefined \def \showDOI       #1{#1}\fi
\ifx \showISBNx    \undefined \def \showISBNx     #1{\unskip}     \fi
\ifx \showISBNxiii \undefined \def \showISBNxiii  #1{\unskip}     \fi
\ifx \showISSN     \undefined \def \showISSN      #1{\unskip}     \fi
\ifx \showLCCN     \undefined \def \showLCCN      #1{\unskip}     \fi
\ifx \shownote     \undefined \def \shownote      #1{#1}          \fi
\ifx \showarticletitle \undefined \def \showarticletitle #1{#1}   \fi
\ifx \showURL      \undefined \def \showURL       {\relax}        \fi
\providecommand\bibfield[2]{#2}
\providecommand\bibinfo[2]{#2}
\providecommand\natexlab[1]{#1}
\providecommand\showeprint[2][]{arXiv:#2}

\bibitem[Alur et~al\mbox{.}(2013)]%
        {AlurAGKV13}
\bibfield{author}{\bibinfo{person}{Rajeev Alur}, \bibinfo{person}{Loris
  D'Antoni}, \bibinfo{person}{Sumit Gulwani}, \bibinfo{person}{Dileep Kini},
  {and} \bibinfo{person}{Mahesh Viswanathan}.} \bibinfo{year}{2013}\natexlab{}.
\newblock \showarticletitle{Automated Grading of {DFA} Constructions}. In
  \bibinfo{booktitle}{\emph{{IJCAI} 2013, Proceedings of the 23rd International
  Joint Conference on Artificial Intelligence}},
  \bibfield{editor}{\bibinfo{person}{Francesca Rossi}} (Ed.).
  \bibinfo{publisher}{{IJCAI/AAAI}}, \bibinfo{pages}{1976--1982}.
\newblock
\urldef\tempurl%
\url{https://www.ijcai.org/Proceedings/13/Papers/292.pdf}
\showURL{%
\tempurl}


\bibitem[Bushmeleva and Baklashova(2017)]%
        {BushmelevaB17}
\bibfield{author}{\bibinfo{person}{Natalya~A. Bushmeleva} {and}
  \bibinfo{person}{Tatiana~A. Baklashova}.} \bibinfo{year}{2017}\natexlab{}.
\newblock \showarticletitle{Methodological Teaching System of Mathematical
  Foundations of Formal Languages as a Means of Fundamentalization of
  Education}.
\newblock \bibinfo{journal}{\emph{EURASIA Journal of Mathematics Science and
  Technology Education}}  \bibinfo{volume}{13} (\bibinfo{year}{2017}),
  \bibinfo{pages}{5141--5155}.
\newblock
Issue 8.
\urldef\tempurl%
\url{https://doi.org/10.12973/eurasia.2017.00989a}
\showDOI{\tempurl}


\bibitem[Creus and Godoy(2014)]%
        {CreusG14}
\bibfield{author}{\bibinfo{person}{Carles Creus} {and} \bibinfo{person}{Guillem
  Godoy}.} \bibinfo{year}{2014}\natexlab{}.
\newblock \showarticletitle{Automatic Evaluation of Context-Free Grammars
  (System Description)}. In \bibinfo{booktitle}{\emph{Rewriting and Typed
  Lambda Calculi}}, \bibfield{editor}{\bibinfo{person}{Gilles Dowek}} (Ed.).
  \bibinfo{publisher}{Springer International Publishing},
  \bibinfo{address}{Cham}, \bibinfo{pages}{139--148}.
\newblock
\showISBNx{978-3-319-08918-8}
\urldef\tempurl%
\url{https://doi.org/10.1007/978-3-319-08918-8_10}
\showDOI{\tempurl}


\bibitem[D'Antoni et~al\mbox{.}(2020)]%
        {AntoniHKRW2020}
\bibfield{author}{\bibinfo{person}{Loris D'Antoni}, \bibinfo{person}{Martin
  Helfrich}, \bibinfo{person}{Jan K\v{r}et\'{i}nsk\'{y}},
  \bibinfo{person}{Emanuel Ramneantu}, {and} \bibinfo{person}{Maximilian
  Weininger}.} \bibinfo{year}{2020}\natexlab{}.
\newblock \showarticletitle{Automata Tutor v3}. In
  \bibinfo{booktitle}{\emph{Computer Aided Verification -- 32nd International
  Conference, {CAV} 2020, Proceedings, Part {II}}}
  \emph{(\bibinfo{series}{Lecture Notes in Computer Science},
  Vol.~\bibinfo{volume}{12225})},
  \bibfield{editor}{\bibinfo{person}{Shuvendu~K. Lahiri} {and}
  \bibinfo{person}{Chao Wang}} (Eds.). \bibinfo{publisher}{Springer},
  \bibinfo{pages}{3--14}.
\newblock
\urldef\tempurl%
\url{https://doi.org/10.1007/978-3-030-53291-8\_1}
\showDOI{\tempurl}


\bibitem[D'Antoni et~al\mbox{.}(2015)]%
        {AntoniKAGV2015}
\bibfield{author}{\bibinfo{person}{Loris D'Antoni}, \bibinfo{person}{Dileep
  Kini}, \bibinfo{person}{Rajeev Alur}, \bibinfo{person}{Sumit Gulwani},
  \bibinfo{person}{Mahesh Viswanathan}, {and} \bibinfo{person}{Bj{\"{o}}rn
  Hartmann}.} \bibinfo{year}{2015}\natexlab{}.
\newblock \showarticletitle{How Can Automatic Feedback Help Students Construct
  Automata?}
\newblock \bibinfo{journal}{\emph{{ACM} Transactions on Computer-Human
  Interaction}} \bibinfo{volume}{22}, \bibinfo{number}{2}
  (\bibinfo{year}{2015}), \bibinfo{pages}{pp.~9:1--9:24}.
\newblock
\urldef\tempurl%
\url{https://doi.org/10.1145/2723163}
\showDOI{\tempurl}


\bibitem[Davis et~al\mbox{.}(1993)]%
        {DavidLMC93}
\bibfield{author}{\bibinfo{person}{E.~A. Davis}, \bibinfo{person}{M.~C. Linn},
  \bibinfo{person}{L.~M. Mann}, {and} \bibinfo{person}{M.~J. Clancy}.}
  \bibinfo{year}{1993}\natexlab{}.
\newblock \showarticletitle{Mind your P's and Q's: Using parentheses and quotes
  in LISP}. In \bibinfo{booktitle}{\emph{Empirical Studies of Programmers:
  Fifth Workshop}}, \bibfield{editor}{\bibinfo{person}{C.~R. Cook},
  \bibinfo{person}{J.~C. Scholtz}, {and} \bibinfo{person}{J.~C. Spohrer}}
  (Eds.). \bibinfo{pages}{63--85}.
\newblock


\bibitem[de~Souza et~al\mbox{.}(2015)]%
        {Souza15}
\bibfield{author}{\bibinfo{person}{Gabriel~Spadon de Souza},
  \bibinfo{person}{Celso Olivete}, \bibinfo{person}{Ronaldo~Celso
  Messias~Correia}, {and} \bibinfo{person}{Rogério~Eduardo Garcia}.}
  \bibinfo{year}{2015}\natexlab{}.
\newblock \showarticletitle{Teaching-learning methodology for formal languages
  and automata theory}. In \bibinfo{booktitle}{\emph{IEEE Frontiers in
  Education Conference (FIE)}}. \bibinfo{pages}{1--7}.
\newblock
\urldef\tempurl%
\url{https://doi.org/10.1109/FIE.2015.7344185}
\showDOI{\tempurl}


\bibitem[Gal-Ezer and Trakhtenbrot(2016)]%
        {GalEzerT16}
\bibfield{author}{\bibinfo{person}{Judith Gal-Ezer} {and} \bibinfo{person}{Mark
  Trakhtenbrot}.} \bibinfo{year}{2016}\natexlab{}.
\newblock \showarticletitle{Identification and addressing reduction-related
  misconceptions}.
\newblock \bibinfo{journal}{\emph{Computer Science Education}}
  \bibinfo{volume}{26}, \bibinfo{number}{2-3} (\bibinfo{year}{2016}),
  \bibinfo{pages}{89--103}.
\newblock
\urldef\tempurl%
\url{https://doi.org/10.1080/08993408.2016.1171470}
\showDOI{\tempurl}


\bibitem[Geck et~al\mbox{.}(2018)]%
        {GeckLPSVZ18}
\bibfield{author}{\bibinfo{person}{Gaetano Geck}, \bibinfo{person}{Artur
  Ljulin}, \bibinfo{person}{Sebastian Peter}, \bibinfo{person}{Jonas Schmidt},
  \bibinfo{person}{Fabian Vehlken}, {and} \bibinfo{person}{Thomas Zeume}.}
  \bibinfo{year}{2018}\natexlab{}.
\newblock \showarticletitle{Introduction to {Iltis}: an interactive, web-based
  system for teaching logic}. In \bibinfo{booktitle}{\emph{Proceedings of the
  23rd Annual {ACM} Conference on Innovation and Technology in Computer Science
  Education, ITiCSE 2018}}. \bibinfo{publisher}{{ACM}},
  \bibinfo{pages}{141--146}.
\newblock
\urldef\tempurl%
\url{https://doi.org/10.1145/3197091.3197095}
\showDOI{\tempurl}


\bibitem[Ginat and Blau(2017)]%
        {GinatB17}
\bibfield{author}{\bibinfo{person}{David Ginat} {and} \bibinfo{person}{Yoav
  Blau}.} \bibinfo{year}{2017}\natexlab{}.
\newblock \showarticletitle{Multiple Levels of Abstraction in Algorithmic
  Problem Solving}. In \bibinfo{booktitle}{\emph{Proceedings of the 2017 ACM
  SIGCSE Technical Symposium on Computer Science Education}} (Seattle,
  Washington, USA) \emph{(\bibinfo{series}{SIGCSE '17})}.
  \bibinfo{publisher}{Association for Computing Machinery},
  \bibinfo{address}{New York, NY, USA}, \bibinfo{pages}{237–242}.
\newblock
\showISBNx{9781450346986}
\urldef\tempurl%
\url{https://doi.org/10.1145/3017680.3017801}
\showDOI{\tempurl}


\bibitem[Gramond and Rodger(1999)]%
        {GramondR1999}
\bibfield{author}{\bibinfo{person}{Eric Gramond} {and}
  \bibinfo{person}{Susan~H. Rodger}.} \bibinfo{year}{1999}\natexlab{}.
\newblock \showarticletitle{Using {JFLAP} to interact with theorems in automata
  theory}. In \bibinfo{booktitle}{\emph{Proceedings of the 30th {SIGCSE}
  Technical Symposium on Computer Science Education, 1999}},
  \bibfield{editor}{\bibinfo{person}{Jane Prey} {and}
  \bibinfo{person}{Robert~E. Noonan}} (Eds.). \bibinfo{publisher}{{ACM}},
  \bibinfo{pages}{336--340}.
\newblock
\urldef\tempurl%
\url{https://doi.org/10.1145/299649.299800}
\showDOI{\tempurl}


\bibitem[Hielscher and Wagenknecht(2019)]%
        {HielscherW19}
\bibfield{author}{\bibinfo{person}{Michael Hielscher} {and}
  \bibinfo{person}{Christian Wagenknecht}.} \bibinfo{year}{2019}\natexlab{}.
\newblock \showarticletitle{{FLACI} -- Eine {L}ernumgebung f{\"{u}}r
  theoretische {I}nformatik}. In \bibinfo{booktitle}{\emph{Informatik f{\"{u}}r
  alle, 18. GI-Fachtagung Informatik und Schule, {INFOS} 2019}}
  \emph{(\bibinfo{series}{{LNI}}, Vol.~\bibinfo{volume}{{P-288}})},
  \bibfield{editor}{\bibinfo{person}{Arno Pasternak}} (Ed.).
  \bibinfo{publisher}{Gesellschaft f{\"{u}}r Informatik},
  \bibinfo{pages}{211--220}.
\newblock
\urldef\tempurl%
\url{https://doi.org/10.18420/infos2019-c6}
\showDOI{\tempurl}


\bibitem[Ishikawa et~al\mbox{.}(2015)]%
        {IshikawaYT15}
\bibfield{author}{\bibinfo{person}{Fuyuki Ishikawa}, \bibinfo{person}{Nobukazu
  Yoshioka}, {and} \bibinfo{person}{Yoshinori Tanabe}.}
  \bibinfo{year}{2015}\natexlab{}.
\newblock \showarticletitle{Keys and Roles of Formal Methods Education for
  Industry: 10 Year Experience with Top SE Program}. In
  \bibinfo{booktitle}{\emph{FMSEE\&T@FM}}.
\newblock
\urldef\tempurl%
\url{https://api.semanticscholar.org/CorpusID:14404471}
\showURL{%
\tempurl}


\bibitem[{Joint Task Force on Computing Curricula, Association for Computing
  Machinery (ACM)} and {IEEE Computer Society}(2013)]%
        {ACM2013}
\bibfield{author}{\bibinfo{person}{{Joint Task Force on Computing Curricula,
  Association for Computing Machinery (ACM)}} {and} \bibinfo{person}{{IEEE
  Computer Society}}.} \bibinfo{year}{2013}\natexlab{}.
\newblock \bibinfo{booktitle}{\emph{Computer Science Curricula 2013: Curriculum
  Guidelines for Undergraduate Degree Programs in Computer Science}}.
\newblock \bibinfo{publisher}{Association for Computing Machinery},
  \bibinfo{address}{New York, NY, USA}.
\newblock
\showISBNx{9781450323093}


\bibitem[Mayring(2014)]%
        {Mayring2014}
\bibfield{author}{\bibinfo{person}{Philipp Mayring}.}
  \bibinfo{year}{2014}\natexlab{}.
\newblock \bibinfo{booktitle}{\emph{Qualitative content analysis: theoretical
  foundation, basic procedures and software solution}}.
\newblock \bibinfo{address}{Klagenfurt}. 143 pages.
\newblock


\bibitem[Mohammed and Shaffer(2024)]%
        {MohammedS24}
\bibfield{author}{\bibinfo{person}{Mostafa Mohammed} {and}
  \bibinfo{person}{Clifford~A. Shaffer}.} \bibinfo{year}{2024}\natexlab{}.
\newblock \showarticletitle{Teaching Formal Languages through Programmed
  Instruction}. In \bibinfo{booktitle}{\emph{Proceedings of the 55th ACM
  Technical Symposium on Computer Science Education V. 1}} (<conf-loc>,
  <city>Portland</city>, <state>OR</state>, <country>USA</country>,
  </conf-loc>) \emph{(\bibinfo{series}{SIGCSE 2024})}.
  \bibinfo{publisher}{Association for Computing Machinery},
  \bibinfo{address}{New York, NY, USA}, \bibinfo{pages}{867–873}.
\newblock
\showISBNx{9798400704239}
\urldef\tempurl%
\url{https://doi.org/10.1145/3626252.3630940}
\showDOI{\tempurl}


\bibitem[Morazán and Antunez(2014)]%
        {MorazanA14}
\bibfield{author}{\bibinfo{person}{Marco~T. Morazán} {and}
  \bibinfo{person}{Rosario Antunez}.} \bibinfo{year}{2014}\natexlab{}.
\newblock \showarticletitle{Functional Automata - Formal Languages for Computer
  Science Students}.
\newblock \bibinfo{journal}{\emph{Electronic Proceedings in Theoretical
  Computer Science}}  \bibinfo{volume}{170} (\bibinfo{year}{2014}),
  \bibinfo{pages}{19–32}.
\newblock
\showISSN{2075-2180}
\urldef\tempurl%
\url{https://doi.org/10.4204/eptcs.170.2}
\showDOI{\tempurl}


\bibitem[Qian et~al\mbox{.}(2020)]%
        {QianHYGL20}
\bibfield{author}{\bibinfo{person}{Yizhou Qian}, \bibinfo{person}{Susanne
  Hambrusch}, \bibinfo{person}{Aman Yadav}, \bibinfo{person}{Sarah Gretter},
  {and} \bibinfo{person}{Yue Li}.} \bibinfo{year}{2020}\natexlab{}.
\newblock \showarticletitle{Teachers' Perceptions of Student Misconceptions in
  Introductory Programming}.
\newblock \bibinfo{journal}{\emph{Journal of Educational Computing Research}}
  \bibinfo{volume}{58}, \bibinfo{number}{2} (\bibinfo{year}{2020}),
  \bibinfo{pages}{364--397}.
\newblock
\urldef\tempurl%
\url{https://doi.org/10.1177/0735633119845413}
\showDOI{\tempurl}


\bibitem[Qian and Lehman(2017)]%
        {Qian17}
\bibfield{author}{\bibinfo{person}{Yizhou Qian} {and} \bibinfo{person}{James
  Lehman}.} \bibinfo{year}{2017}\natexlab{}.
\newblock \showarticletitle{Students' Misconceptions and Other Difficulties in
  Introductory Programming: A Literature Review}.
\newblock \bibinfo{journal}{\emph{ACM Trans. Comput. Educ.}}
  \bibinfo{volume}{18}, \bibinfo{number}{1}, Article \bibinfo{articleno}{1}
  (\bibinfo{date}{oct} \bibinfo{year}{2017}), \bibinfo{numpages}{24}~pages.
\newblock
\urldef\tempurl%
\url{https://doi.org/10.1145/3077618}
\showDOI{\tempurl}


\bibitem[Renée~McCauley and Murphy(2015)]%
        {McCauleyGFM15}
\bibfield{author}{\bibinfo{person}{Sue~Fitzgerald Renée~McCauley,
  Scott~Grissom} {and} \bibinfo{person}{Laurie Murphy}.}
  \bibinfo{year}{2015}\natexlab{}.
\newblock \showarticletitle{Teaching and learning recursive programming: a
  review of the research literature}.
\newblock \bibinfo{journal}{\emph{Computer Science Education}}
  \bibinfo{volume}{25}, \bibinfo{number}{1} (\bibinfo{year}{2015}),
  \bibinfo{pages}{37--66}.
\newblock
\urldef\tempurl%
\url{https://doi.org/10.1080/08993408.2015.1033205}
\showDOI{\tempurl}


\bibitem[Rodger(1999)]%
        {Rodger1999}
\bibfield{author}{\bibinfo{person}{Susan~H. Rodger}.}
  \bibinfo{year}{1999}\natexlab{}.
\newblock \showarticletitle{Teaching automata theory with {JFLAP}}.
\newblock \bibinfo{journal}{\emph{{SIGACT} News}} \bibinfo{volume}{30},
  \bibinfo{number}{4} (\bibinfo{year}{1999}), \bibinfo{pages}{53--56}.
\newblock
\urldef\tempurl%
\url{https://doi.org/10.1145/337885.337896}
\showDOI{\tempurl}


\bibitem[Scheurer(2000)]%
        {Scheurer00}
\bibfield{author}{\bibinfo{person}{Thierry Scheurer}.}
  \bibinfo{year}{2000}\natexlab{}.
\newblock \showarticletitle{Formal Methods: The Problem Is Education}. In
  \bibinfo{booktitle}{\emph{Computer Safety, Reliability and Security}},
  \bibfield{editor}{\bibinfo{person}{Floor Koornneef} {and}
  \bibinfo{person}{Meine van~der Meulen}} (Eds.). \bibinfo{publisher}{Springer
  Berlin Heidelberg}, \bibinfo{address}{Berlin, Heidelberg},
  \bibinfo{pages}{198--210}.
\newblock
\showISBNx{978-3-540-40891-8}
\urldef\tempurl%
\url{https://doi.org/10.1007/3-540-40891-6_18}
\showDOI{\tempurl}


\bibitem[Schmellenkamp et~al\mbox{.}(2023)]%
        {SchmellenkampLZ23}
\bibfield{author}{\bibinfo{person}{Marko Schmellenkamp},
  \bibinfo{person}{Alexandra Latys}, {and} \bibinfo{person}{Thomas Zeume}.}
  \bibinfo{year}{2023}\natexlab{}.
\newblock \showarticletitle{Discovering and Quantifying Misconceptions in
  Formal Methods Using Intelligent Tutoring Systems}. In
  \bibinfo{booktitle}{\emph{Proceedings of the 54th ACM Technical Symposium on
  Computer Science Education V. 1}} (Toronto ON, Canada)
  \emph{(\bibinfo{series}{SIGCSE 2023})}. \bibinfo{publisher}{Association for
  Computing Machinery}, \bibinfo{address}{New York, NY, USA},
  \bibinfo{pages}{465–471}.
\newblock
\showISBNx{9781450394314}
\urldef\tempurl%
\url{https://doi.org/10.1145/3545945.3569806}
\showDOI{\tempurl}


\bibitem[Schmellenkamp et~al\mbox{.}(2024)]%
        {SchmellenkampVZ24}
\bibfield{author}{\bibinfo{person}{Marko Schmellenkamp},
  \bibinfo{person}{Fabian Vehlken}, {and} \bibinfo{person}{Thomas Zeume}.}
  \bibinfo{year}{2024}\natexlab{}.
\newblock \showarticletitle{Teaching Formal Foundations of Computer Science
  with {Iltis}}.
\newblock \bibinfo{journal}{\emph{Bulleting of {EATCS}}}  \bibinfo{volume}{142}
  (\bibinfo{year}{2024}).
\newblock


\bibitem[Sfard(1995)]%
        {Sfard95}
\bibfield{author}{\bibinfo{person}{Anna Sfard}.}
  \bibinfo{year}{1995}\natexlab{}.
\newblock \showarticletitle{The development of algebra: Confronting historical
  and psychological perspectives}.
\newblock \bibinfo{journal}{\emph{The Journal of Mathematical Behavior}}
  \bibinfo{volume}{14}, \bibinfo{number}{1} (\bibinfo{year}{1995}),
  \bibinfo{pages}{15--39}.
\newblock
\showISSN{0732-3123}
\urldef\tempurl%
\url{https://doi.org/10.1016/0732-3123(95)90022-5}
\showDOI{\tempurl}


\bibitem[Sigman(2007)]%
        {Sigman07}
\bibfield{author}{\bibinfo{person}{Scott Sigman}.}
  \bibinfo{year}{2007}\natexlab{}.
\newblock \showarticletitle{Engaging students in formal language theory and
  theory of computation}.
\newblock \bibinfo{journal}{\emph{SIGCSE Bull.}} \bibinfo{volume}{39},
  \bibinfo{number}{1} (\bibinfo{date}{mar} \bibinfo{year}{2007}),
  \bibinfo{pages}{450–453}.
\newblock
\showISSN{0097-8418}
\urldef\tempurl%
\url{https://doi.org/10.1145/1227504.1227463}
\showDOI{\tempurl}


\bibitem[Spichkova and Zamansky(2016)]%
        {SpichkovaZ16}
\bibfield{author}{\bibinfo{person}{Maria Spichkova} {and} \bibinfo{person}{Anna
  Zamansky}.} \bibinfo{year}{2016}\natexlab{}.
\newblock \showarticletitle{Teaching of Formal Methods for Software
  Engineering}. In \bibinfo{booktitle}{\emph{Proceedings of the 11th
  International Conference on Evaluation of Novel Software Approaches to
  Software Engineering - Volume 1: COLAFORM, (ENASE 2016)}}. INSTICC,
  \bibinfo{publisher}{SciTePress}, \bibinfo{pages}{370--376}.
\newblock
\showISBNx{978-989-758-189-2}
\urldef\tempurl%
\url{https://doi.org/10.5220/0005928503700376}
\showDOI{\tempurl}


\bibitem[Swidan et~al\mbox{.}(2018)]%
        {SwidanHS18}
\bibfield{author}{\bibinfo{person}{Alaaeddin Swidan}, \bibinfo{person}{Felienne
  Hermans}, {and} \bibinfo{person}{Marileen Smit}.}
  \bibinfo{year}{2018}\natexlab{}.
\newblock \showarticletitle{Programming Misconceptions for School Students}. In
  \bibinfo{booktitle}{\emph{Proceedings of the 2018 ACM Conference on
  International Computing Education Research}} (Espoo, Finland)
  \emph{(\bibinfo{series}{ICER '18})}. \bibinfo{publisher}{Association for
  Computing Machinery}, \bibinfo{address}{New York, NY, USA},
  \bibinfo{pages}{151–159}.
\newblock
\showISBNx{9781450356282}
\urldef\tempurl%
\url{https://doi.org/10.1145/3230977.3230995}
\showDOI{\tempurl}


\bibitem[Tavolato and Vogt(2012)]%
        {TavolatoV12}
\bibfield{author}{\bibinfo{person}{Paul Tavolato} {and}
  \bibinfo{person}{Friedrich Vogt}.} \bibinfo{year}{2012}\natexlab{}.
\newblock \showarticletitle{Integrating formal methods into computer science
  curricula at a university of applied sciences}. In
  \bibinfo{booktitle}{\emph{TLA+ workshop at the 18th international symposium
  on formal methods}}.
\newblock


\bibitem[Tscherter(2004)]%
        {Tscherter04}
\bibfield{author}{\bibinfo{person}{Vincent Tscherter}.}
  \bibinfo{year}{2004}\natexlab{}.
\newblock \emph{\bibinfo{title}{Exorciser. Automatic generation and interactive
  grading of structured excercises in the theory of computation}}.
\newblock Doctoral Thesis. \bibinfo{school}{ETH Zurich},
  \bibinfo{address}{Zürich}.
\newblock
\urldef\tempurl%
\url{https://doi.org/10.3929/ethz-a-004830877}
\showDOI{\tempurl}
\newblock
\shownote{Diss., Technische Wissenschaften ETH Zürich, Nr. 15654, 2004.}.


\bibitem[Webb et~al\mbox{.}(2017)]%
        {Webb17}
\bibfield{author}{\bibinfo{person}{Mary Webb}, \bibinfo{person}{Niki Davis},
  \bibinfo{person}{Tim Bell}, \bibinfo{person}{Yaacov~J Katz},
  \bibinfo{person}{Nicholas Reynolds}, \bibinfo{person}{Dianne~P Chambers},
  {and} \bibinfo{person}{Maciej~M Sys{\l}o}.} \bibinfo{year}{2017}\natexlab{}.
\newblock \showarticletitle{Computer science in K-12 school curricula of the
  21st century: Why, what and when?}
\newblock \bibinfo{journal}{\emph{Education and Information Technologies}}
  \bibinfo{volume}{22} (\bibinfo{year}{2017}), \bibinfo{pages}{445--468}.
\newblock


\bibitem[Witzel et~al\mbox{.}(2008)]%
        {WitzelRS08}
\bibfield{author}{\bibinfo{person}{Bradley~S. Witzel}, \bibinfo{person}{Paul~J.
  Riccomini}, {and} \bibinfo{person}{Elke Schneider}.}
  \bibinfo{year}{2008}\natexlab{}.
\newblock \showarticletitle{Implementing CRA With Secondary Students With
  Learning Disabilities in Mathematics}.
\newblock \bibinfo{journal}{\emph{Intervention in School and Clinic}}
  \bibinfo{volume}{43}, \bibinfo{number}{5} (\bibinfo{year}{2008}),
  \bibinfo{pages}{270--276}.
\newblock
\urldef\tempurl%
\url{https://doi.org/10.1177/1053451208314734}
\showDOI{\tempurl}


\bibitem[Young and O'Shea(1981)]%
        {YoungO81}
\bibfield{author}{\bibinfo{person}{Richard~M. Young} {and} \bibinfo{person}{Tim
  O'Shea}.} \bibinfo{year}{1981}\natexlab{}.
\newblock \showarticletitle{Errors in Children's Subtraction}.
\newblock \bibinfo{journal}{\emph{Cognitive Science}} \bibinfo{volume}{5},
  \bibinfo{number}{2} (\bibinfo{year}{1981}), \bibinfo{pages}{153--177}.
\newblock
\urldef\tempurl%
\url{https://doi.org/10.1207/s15516709cog0502_3}
\showDOI{\tempurl}


\end{thebibliography}

\end{document}